\documentclass[11pt]{article}

\usepackage[all]{xy}
\usepackage{comment,cite,amsmath,amssymb}

\newcommand{\ba}{\begin{eqnarray}}
\newcommand{\ea}{\end{eqnarray}}
\newcommand{\nn}{\nonumber}
\newcommand{\nt}{\nonumber\\}
\newcommand{\vm}{{\vec m}}
\newcommand{\vn}{{\vec n}}

\newcommand{\ud}[3]{{#1}^{#2}_{\phantom{#2}#3}}
\newcommand{\cN}{{\cal N}}

\newcommand{\ha}{{\hat a}}
\newcommand{\hb}{{\hat b}}
\newcommand{\hc}{{\hat c}}
\newcommand{\hd}{{\hat d}}
\newcommand{\he}{{\hat e}}
\newcommand{\hf}{{\hat f}}
\newcommand{\hg}{{\hat g}}
\newcommand{\hh}{{\hat h}}
\newcommand{\bR}{{\bf R}}

\begin{document}
\begin{titlepage}

\begin{center}

\hfill UT-09-11
\vskip .5in

\textbf{\LARGE Aspects of U-duality in BLG models\\[+2pt]
with Lorentzian metric 3-algebras}

\vskip .5in
{\large
Takayuki Kobo\footnote{
E-mail address: kobo@hep-th.phys.s.u-tokyo.ac.jp},
Yutaka Matsuo\footnote{
E-mail address: matsuo@phys.s.u-tokyo.ac.jp},
Shotaro Shiba\footnote{
E-mail address: shiba@hep-th.phys.s.u-tokyo.ac.jp}
}\\
\vskip 10mm
{\it\large
Department of Physics, Faculty of Science, University of Tokyo,\\
Hongo 7-3-1, Bunkyo-ku, Tokyo 113-0033, Japan\\
\noindent{ \smallskip }\\
}
\vspace{60pt}
\end{center}
\begin{abstract}
In \cite{HMS}, it was shown that BLG model based
on a Lorentzian metric 3-algebra
 gives D$p$-brane action whose worldvolume
is compactified on torus $T^{d}$ ($d=p-2$).  
Here the 3-algebra was a generalized one
with $d+1$ pairs of Lorentzian metric generators
and expressed in terms of
 a loop algebra with central extensions. 
In this paper, we derive the precise relation
between the coupling constant of the super Yang-Mills, the 
moduli of $T^d$ and some R-R flux with VEV's
of ghost fields associated with Lorentzian metric generators. 
In particular, for $d=1$, we derive the Yang-Mills action
with $\theta$ term and show that $SL(2,\mathbf{Z})$ Montonen-Olive
duality is realized as the rotation of two VEV's.
Furthermore, some moduli parameters such as NS-NS 2-form flux
are identified as the deformation parameters of 
the 3-algebras.  By combining them, we recover most of
the moduli parameters which are required by U-duality symmetry.
\end{abstract}

\end{titlepage}

\setcounter{footnote}{0}

\section{Introduction and Summary}
Recently, Bagger, Lambert \cite{BL} and Gustavsson \cite{G} found 
that a certain class of Chern-Simons matter system can have 
maximal supersymmetry in $2+1$ dimensions
and that it may describe the multiple M2-branes.  Their action is
distinctive in that the gauge symmetry is based on a new
mathematical framework, Lie 3-algebra.
However,  it was soon realized that their constraints on the algebra are too 
restrictive that the only allowed 3-algebra is so-called ${\cal A}_4$ 
algebra which describes the two M2-branes~\cite{no-go}. 

For the description of larger number of M2-branes, 
many study have been made to generalize the BLG framework.%
\footnote{{}Apart from the examples mentioned below, there is also
an example based on the 3-algebra with Nambu-Poisson bracket~\cite{M5}.
This algebra describes the infinite number of M2-branes and realizes 
the worldvolume theory of a single M5-brane in the $C$-field background 
on a 3-manifold where Nambu-Poisson bracket is equipped.}
The first interesting example was found by three groups 
\cite{HIM,GMR,Benvenuti:2008bt}
which is based on the 3-algebra with a pair of Lorentzian metric generators
$u, v$ and arbitrary Lie algebra generators $T^i$, such that
\ba
&&[u, T^i, T^j]= i {f^{ij}}_k T^k\,,\quad
[T^i, T^j, T^k]= -if^{ijk} v\,,\nt
&&\langle u,v\rangle=1\,,\quad
\langle T^i,T^j\rangle=\delta^{ij}\,,\label{LL3}
\ea
where we keep only the nonvanishing 3-commutators and metric components.
While the components associated with the generators $u,v$ become ghosts,
they can be removed by a new kind of Higgs mechanism proposed by
\cite{MP}. After the ghost is removed, the Chern-Simons matter system
is reduced to the ordinary super Yang-Mills system which describes 
multiple D2-branes. Then many studies are undertaken on this Lorentzian 
BLG model~\cite{rLBLG,HMS,dMFM}. However, since the correspondence is too exact, 
the model was realized to be too simple to describe the full M2-brane dynamics.

Soon after, another $2+1$ dimensional Chern-Simons matter system 
with $SU(N)\times SU(N)$ gauge symmetry was proposed \cite{ABJM}.  
While it lacks the manifest $\cN=8$ supersymmetry,
it has many attractive features such as the brane construction,
AdS/CFT correspondence, and relation with the integrable spin chain.
In particular, it gives a good description of M2-branes when
the coupling constant $N/k$ ($k$ is the level of Chern-Simons term)
becomes small.

The models based on the Lorentzian
metric 3-algebras  \cite{HIM,GMR,Benvenuti:2008bt}
which was later generalized in~\cite{HMS,dMFM}
by including more Lorentzian metric generators
(in the following, we call it `L-BLG model' 
in short), nevertheless,
still enjoy unique advantages that they keep $\cN=8$ supersymmetry 
as well as $SO(8)$ R-symmetry. 
Of course, M-theory requires 
such symmetry explicitly, so we believe that 
L-BLG models will be able to provide some 
nontrivial information on M-theory.

In this paper, as one of such examples, we examine how U-duality \cite{U-duality}
is realized in L-BLG models.%
\footnote{The Montonen-Olive duality in ABJM context was discussed in \cite{HTT}.
In their study, the coupling constants of the super Yang-Mills
are restricted to depend only one real variable.
In our case, there is no such limitation.}
It is based on a work \cite{HMS} where
a description of M-theory on higher dimensional torus $T^{d+1}$
was given by generalization of 3-algebra with
more Lorentzian metric pairs, say $(u_A, v^A)$
($A=0,1,\cdots, d$).\footnote{
Somewhat similar analysis was made on the generalization
of the Lorentzian metric  \cite{dMFM}.
Their analysis was limited to the finite dimensional cases
and does not include the 3-algebra which is the main focus
of this paper.
}
As a generalization of
the original model, we have $d+1$ pairs of the ghost fields
associated with each $(u_A, v^A)$.
By choosing the structure of 3-algebra carefully, 
it has been shown that such ghost modes can be removed
and the system becomes unitary as in the original model.
In this Higgs mechanism, one has to assign VEV's to these ghost fields as
\ba
X^I_{u_A}= \lambda^{IA}\,,\quad
\vec \lambda^A\in \mathbf{R}^{d+1}\subset \mathbf{R}^8\,.
\ea
These VEV's $\vec\lambda^A$, in turn, describe how 
the transverse directions $\mathbf{R}^8$ are compactified on $T^{d+1}$.
In other words, the Higgs mechanism of L-BLG model produces
the Kaluza-Klein mass associated such compactification.
In \cite{HMS}, it was shown that L-BLG model
gives a super Yang-Mills system whose worldvolume
is a flat  $T^d$ bundle on $\mathcal{M}$, 
where $\mathcal{M}$ is the worldvolume of BLG model.  
In the section 2 of this paper,
we perform a more detailed analysis with general $\vec \lambda^A$ and
determine the precise relation between 
the coupling constant, moduli of the torus $T^d$,
and some R-R flux 
on D$p$-brane worldvolume theory 
in terms of VEV's $\vec \lambda^A$ of L-BLG model.

These parameters are sufficient to fix all the moduli of 
D3-branes theory that corresponds to $d=1$ case.  
Indeed, in the section 3, we argue that the action thus derived
reproduces the complete 4-dimensional super Yang-Mills action with  
$\theta$ term. In particular, Montonen-Olive
$SL(2,\mathbf{Z})$ duality \cite{Montonen:1977sn} is realized by the rotation of the VEV's,
\ba\label{rotVEV}
\vec\lambda'^A= {\ud \Lambda A B} \vec \lambda^B\,,\quad
{\ud \Lambda A B} \in SL(2,\mathbf{Z})\,.
\ea 
While we do not claim that we prove the duality symmetry,
the simplicity of the realization is nevertheless remarkable.
For $d>1$, it is natural to guess that 
the $SL(d+1,\mathbf{Z})$ part of the U-duality
transformation 
is described by the change of the basis
as (\ref{rotVEV}) where $\Lambda\in SL(d+1,\mathbf{Z})$.
We note that U-duality group is give by a product
 $SL(d+1;\mathbf{Z})\bowtie
O(d,d;\mathbf{Z})=:E_{d+1 (d+1)}(\mathbf{Z})$, where the symbol $\bowtie$ denotes the group 
generated by the two non-commuting subgroups
(see, for example, a review article  \cite{Obers:1998fb}).
The $O(d,d;\mathbf{Z})$ part represents the T-duality
symmetry. In our formulation, it is realized by the T-duality relation 
by Taylor \cite{Taylor:1996ik}.

Actually, for $d>1$,
the moduli parameters obtained from Higgs VEV's $\vec\lambda^A$ 
are not enough to realize full U-duality group.
The description of U-duality covariant parameters
for super Yang-Mills system is given in
the context of BFSS matrix theory \cite{Hull, U-mat}.
One of such missing parameters is the NS-NS 2-form flux.
We know already that this parameter can be included
in the theory by the redefinition of the 3-algebra \cite{HMS}.
As $d$ getting larger, we need more R-R flux also.
We give some argument that these extra parameters
will be obtained by changing 3-algebra further, possibly
by including contributions of Nambu-Poisson algebra
as \cite{M5}.

\section{D$p$-brane action from BLG model with moduli parameters}
\label{sec:Dp}

In this section, we perform more detailed  analysis of L-BLG model
which is described in \S5 of \cite{HMS}.
The novelty of the following analysis is to introduce
general VEV's for the ghost fields which gives rise to
the nontrivial metric for the torus $T^d$ and an
extra coupling constants which are related with some R-R flux
on D$p$-brane. The action after Higgs mechanism
is summarized in \S\ref{s:summary}.
We give also more careful explanation of the compactification
mechanism and the geometry of the D$p$-brane worldvolume.

\subsection{BLG Lagrangian and 3-algebra for D$p$-brane}

The original BLG action is written as~\cite{BL}
\ba
S&=&\int_\mathcal{M} d^3x\, L 
=\int_\mathcal{M} d^3x\, (L_X+L_\Psi+L_{int}+L_{pot}+L_{CS}), \label{S} \\
L_X&=&-\frac{1}{2}\langle D_\mu X^I , D^\mu X^I \rangle, \label{LX}\\
L_\Psi&=&\frac{i}{2}\langle \bar\Psi , \Gamma^\mu D_\mu \Psi \rangle, 
\label{LPsi} \\
L_{int}&=&\frac{i}{4}\langle\bar\Psi , \Gamma_{IJ}[X^I,X^J,\Psi] \rangle, 
\label{Lint} \\
L_{pot}&=&
-\frac{1}{12} \langle [X^I, X^J, X^K], [X^I, X^J, X^K]\rangle, 
\label{Lpot} \\
L_{CS}&=& \frac{1}{2}f^{ABCD} A_{AB}\wedge d A_{CD}+
\frac{i}{3} {f^{CDA}}_G f^{EFGB} A_{AB}\wedge
A_{CD}\wedge  A_{EF}\,, \label{LCS}
\ea
where the indices $\mu=0,1,2$ specify the longitudinal directions of M2-branes,
$I,J,K=3,\cdots,10$ indicate the transverse directions,
and the indices $A,B,C,\cdots$ denote components of 3-algebra generators. 
$\mathcal{M}$ is the worldvolume of M2-brane.

The covariant derivative is
\ba \label{eq:cov}
 (D_\mu \Phi(x))_A =\partial_\mu \Phi_A+{f^{CDB}}_A A_{\mu CD}(x) \Phi_B
\ea
for $\Phi=X^I,\Psi$. The 3-bracket for the 3-algebra in BLG model
\ba
{}[T^A,T^B,T^C]=i{\ud f{ABC}D}T^D
\ea
must satisfy the fundamental identity and the invariant metric condition.
Note that the notation is slightly different from the original BLG's one 
in order to make the field $A_{\mu AB}$ Hermite.

In \cite{HMS}, we made a systematic study of Lorentzian
metric  3-algebra which
contains $d+1$ pairs of Lorentzian metric generators $(u_a, v^a)$
together with positive-definite generators $e^i$.
We studied a special class of 3-algebra where the generators $v^a$ 
is the center of 3-algebra, namely $[v^a,\star,\star]=0$,
and the generators $u_a$ are not produced by the any 3-commutators,
{\em i.e.} ${\ud f{\star,\star,\star}{u_a}}=0$.
These requirements are necessary if we want to remove the ghost fields
by using the Higgs mechanism in \cite{MP, HIM}.  A general feature
for $d\geq 1$ is that the gauge fields (as well as all other fields,
because of supersymmetry) become massive by absorbing 
two Higgs (ghost) scalar fields.

For finite dimensional 3-algebras, it is not obvious how to interpret
these massive fields in the context of M/string theory.
It was also found that BLG model based on known finite dimensional
3-algebras produce either products of the supersymmetric gauge 
theories~\cite{HIM,GMR,Benvenuti:2008bt} or abelian massive
super Yang-Mills systems without interactions~\cite{HMS,dMFM}.

For infinite dimensional case, it was found that there are varieties 
of possible 3-algebras and the BLG model associated with them in general
have natural interpretation in M/string theory~\cite{HMS}. 
For example, while the number of particles becomes infinite, 
they are naturally interpreted as the Kaluza-Klein modes
associated with the toroidal compactification.
Also, the mass generated by ghosts can be identified with the Kaluza-Klein mass.

Here we pick up a 3-algebra which produces the
worldvolume theory of D$p$-brane ($p=d+2$): 
\footnote{\label{f:a}
In \cite{HMS}, more general 3-algebra
is considered with the anti-symmetric tensor $C_{ab}$, {\em i.e.}
$[u_0,u_a,u_b]=C_{ab}T^0_{\vec 0}$ instead of eq.\,(\ref{dpa}).
This tensor is related with the noncommutativity parameter
on D$p$-brane.  In this paper, we omit this factor for the
simplicity of the argument.} 
\ba\label{dpa}
&& [u_0, u_a, u_b]=0\,,\\
&& [u_0, u_a, T^i_{\vec m} ]= m_a T^i_{\vec m}\,,\\
&& [u_0, T^i_{\vec m}, T^j_{\vec n} ]= m_a v^a \delta_{\vec m+\vec n} \delta^{ij} 
+ i{f^{ij}}_k T^k_{\vec m+\vec n}\,,\\
&&[T^i_{\vec l}, T^j_{\vec m}, T^k_{\vec n}] = - if^{ijk} \delta_{\vec l+\vec m+\vec n} v^0\,.\label{dpaf}
\ea
where $a,b=1,\cdots, d$\,, $\vec l,\vec m,\vec n\in \mathbf{Z}^d$ and
$f^{ijk}$ ($i,j,k=1,\cdots, \mathrm{dim}\, \mathbf{g}$) is a 
structure constant of an arbitrary Lie algebra $\mathbf{g}$
which satisfies Jacobi identity.
Other 3-commutators are defined to be zero. 
The 3-algebra satisfies the fundamental identity. 
We note that $v^A$ ($A=0,1,\cdots,d$) are
the center of the 3-algebra and $u_A$ do not
appear in the output of 3-commutators.
This is an essential property of Lorentzian metric 3-algebra
to make ghosts disappear after the Higgs mechanism.
The nonvanishing  part of the metric is given as
\ba
\langle u_A, v^B\rangle=\delta_A^{B},\quad
\langle T^i_{\vec m}, T^j_{\vec n}\rangle = \delta^{ij} \delta_{\vec m+\vec n}\,.
\ea

We note that this 3-algebra can be regarded as 
original Lorentzian metric 3-algebra (\ref{LL3})
where Lie algebra is replaced by
\ba \label{la}
&& [u_a, u_b]=0,\quad
[u_a, T^i_{\vec m}]=m_a T^i_{\vec m},\nn\\
&&[T^i_{\vec m}, T^j_{\vec n}]=m_a v^a \delta_{\vec m+\vec n}\delta^{ij}
+i{f^{ij}}_k T^k_{\vec m+\vec n}\,.
\ea
For $d=1$, this is the standard Kac-Moody algebra with
degree operator $u$ and the central charge $v$
and above algebra is its higher loop generalization.
Since the original L-BLG model reduces to super Yang-Mills,
one might guess that BLG model based on the 3-algebra (\ref{dpa}--\ref{dpaf})
should be equivalent to super Yang-Mills whose gauge group is
the loop algebra (\ref{la}).\footnote{We note that the super Yang-Mills system
with loop algebra symmetry is given in \S5.1 of \cite{HMS}.}
It turns out that this is not the case. As we explain below,
BLG Lagrangian contains extra topological terms which
can not be reproduced from Yang-Mills action.

\subsection{Component Expansion}

In the remainder of this section, we will derive the BLG action for 
this 3-algebra.
This was already presented in \cite{HMS} but the computation
is limited to the simplest choice of parameters and the dependence
on the moduli parameter was not clarified. In particular, we will obtain
some ``topological" terms such as $\theta\int F\tilde F$ for D3-brane
which could not show up for the simplest choice of the background.
Furthermore, in order to obtain this $\theta$ term, we must carefully 
deal with the total derivative terms which is neglected in \cite{HMS}.

For the 3-algebra (\ref{dpa}--\ref{dpaf}), we expand various fields as
\ba
X^I &=& X^I_{(i\vec m)} T^i_{\vec m} + X^{IA} u_A +\underline{X}^I_A v^A\\
\Psi &=& \Psi_{(i\vec m)} T^i_{\vec m} + \Psi^A u_A +\underline{\Psi}_A v^A\\
A_\mu &=& A_{\mu (i\vm)(j\vn)}T^i_\vm \wedge T^j_\vn
+ \frac{1}{2} A_{\mu (i\vec m)} u_0 \wedge T^i_{\vec m}
+ \frac12 A_{\mu (i\vm)}^a  u_a \wedge T^i_\vm \nt&&
+ \frac12 A_{\mu}^a \, u_0\wedge u_a 
+ A_{\mu}^{ab}\, u_a\wedge u_b 
+ (\mbox{terms including $v^A$})\,.
\ea
Now we will rewrite the BLG action (\ref{S}) as an action for D$p$-branes 
($p=d+2$).
More precisely, if we denote the original membrane worldvolume
as $\mathcal{M}$, the worldvolume of D$p$-brane is 
given by a flat $T^d$ bundle over $\mathcal{M}$. 
The index $\vec m\in \mathbf{Z}^{d}$ which appears
in some components represents the Kaluza-Klein momentum
along the $T^d$.

In this geometrical set-up, each bosonic components plays the following
roles:
\begin{itemize}
\item $X^I_{(i\vec m)}$\,: These are splitted into three groups.
Some are the collective coordinates which describe the embedding into 
the transverse directions, others are the gauge fields on the worldvolume,
and the other is the degree of freedom which can be absorbed 
when M-direction disappears. The concrete expression is 
eq.\,(\ref{eq:expandX}).
\item $X^{IA}$\,: Higgs fields whose VEV's determine either
the moduli of $T^d$ or the compactification radius in M-direction.
\item $A_{\mu (i\vec m)}$\,: gauge fields along the membrane worldvolume 
$\mathcal{M}$.
\item $A_{\mu}^a$\,: a connection which describes the fiber bundle
$T^d\rightarrow \mathcal{M}$.  The equation of motion
implies that it is always flat $\partial_{[\mu} A_{\nu]}^a=0$.
\end{itemize}
The other bosonic components become Lagrange multiplier or
do not show up in the action at all.
In the following, we set $A_{\mu}^a=A_{\mu}^{ab}=0$ for simplicity.


\subsection{Solving the ghost sector}

The components of ghost fields $\underline X$ and $\underline \Psi$ 
appear in the action only through the following terms:
\ba
L_{gh}=-(D_\mu X^I)_{u_A} (D_\mu X^I)_{v^A}
+\frac{i}{2}\left(\bar\Psi_{u_A}\Gamma^\mu D_\mu\Psi_{v^A}
 +\bar\Psi_{v^A}\Gamma^\mu D_\mu\Psi_{u_A}\right)
\ea
where
\ba
(D_\mu X^I)_{u_A}&=&\partial_\mu X^{IA}\,,\nt
(D_\mu X^I)_{v^0}&=&\partial_\mu \underline{X}^I_0
+im_a(A_{\mu(i\vm)}^a X_{(i,-\vm)}^I+A_{\mu (i\vm)(i,-\vm)}X^{Ia})\nt&&
-f^{ijk}A_{\mu (i\vm)(j\vm)}X_{(k,-\vm-\vn)}^I\,,\nt
(D_\mu X^I)_{v^a}&=&\partial_\mu \underline{X}^I_a
-im_a(A_{\mu (i\vm)}X_{(i,-\vm)}^I+A_{\mu(i\vm)(i,-\vm)}X^{I0})\,,
\ea
and similar for $\Psi$. 
The variation of $\underline{X}^I_A$ and $\underline{\Psi}_A$ always 
give the {\em free} equations of motion for $X^{IA}$ and $\Psi^A$, namely 
\ba
\partial^\mu \partial_\mu X^{IA}=0\,,\quad
\Gamma^\mu\partial_\mu \Psi^A=0\,.
\ea
By introducing extra gauge fields $C^I_{\mu A}$ and $\chi_A$ 
through~\cite{Bandres:2008kj,Gomis:2008be}
\ba
L_{new}={C^I_{\mu A}}\partial_\mu X^{IA}-\chi_A\bar\Psi^A \,,
\ea
one may modify the equations of motion for $X^{IA}$ and $\Psi^A$ to
\ba \label{eqom}
\partial_\mu X^{IA}=0\,,\quad \Psi^A=0\,,
\ea
and absorb the ghosts $\underline{X}^I_A$ and $\underline{\Psi}_A$ 
by gauge fixing. This is how the ghost fields can be removed in \cite{HIM,GMR,Benvenuti:2008bt}.


The equations of motion for $X^{IA}$ (\ref{eqom}) imply 
that they are constant vectors in $\mathbf{R}^8$. 
We fix these constants as
\ba
\vec X^A=\vec \lambda^A\in \mathbf{R}^{d+1}\subset \mathbf{R}^8\,.
\ea
In \cite{HIM,GMR,Benvenuti:2008bt}, there is only one
$\vec\lambda=\vec \lambda^0$ which specifies the M-direction
compactification radius.  
This time, we have extra VEV's $\vec\lambda^a$ which
give the moduli of the toroidal compactification $T^d$.

In the following, we prepare some notations for the later discussion.
We write the dual basis to $\vec \lambda^A$ as $\vec \pi_A$,
which satisfy
\ba
\vec\lambda^A\cdot \vec \pi_B=\delta^A_B \,.
\ea
We introduce a projector into the subspace of $\mathbf{R}^8$ which is orthogonal
to all $\vec \lambda^A$ as
\ba
P^{IJ} = \delta^{IJ}-\sum_A \lambda^{IA}\pi^J_A \,,
\ea
which satisfies $P^2=P$. We define `metric' as
\ba
G^{AB}=\vec\lambda^A\cdot \vec\lambda^B \,,
\ea
where $\lambda^{IA}$ play the role of vierbein.
Using this metric, $\vec \pi_0$ can be written as
\ba \label{eq:pi0}
\vec\pi_0=\frac{1}{G^{00}}\,\vec\lambda^0
-\frac{G^{0a}}{G^{00}}\,\vec\pi_a\,,
\ea
and from now we use $\{\vec\lambda^0,\vec\pi_a\}$ 
as the basis of ${\bf R}^{d+1}$ spanned by $\vec\lambda^A$.
Note that $\vec\lambda^0\perp\vec\pi_a$ for all $a$.
Our claim that the ${\bf R}^{d+1}$ is compactified on $T^{d+1}$
will be deduced from the Kaluza-Klein mass which is generated by 
the Higgs mechanism. This will be demonstrated below.

\paragraph{Comments on Higgs potential}
Since $\vec X^A$ plays the role of Higgs fields, it is natural to 
wonder if one may introduce a potential for them and fix the value
of  VEV's.  This seems to be physically relevant
 since they are related to the moduli of torus.  One naive guess is
 to add a potential $-V(\vec X^A)$ to the action.  Since
the SUSY and gauge transformations of  $\vec X^A$ are trivial,
this potential breaks neither SUSY nor gauge symmetry.
However, the kinetic term is given in the mixed form 
$\partial \vec X^A \partial \underline{\vec X}_A$,
the potential does not fix $\vec X^A$ but 
physically irrelevant $\underline{\vec X}_A$.

\subsection{Derivation of D$p$-brane action}
We finally rewrite the BLG action (\ref{S}) in terms of 3-algebra components
and by putting VEV's to ghost fields $X^{IA}$ and $\Psi^A$.

\paragraph{Kinetic terms for $X^I$ and $\Psi$}
The covariant derivative becomes, after the assignment of VEV's to ghosts,
\ba\label{cvd}
(D_\mu X^I)_{(i\vec m)}
= (\hat D_\mu X^I)_{(i\vec m)}
+ A'_{\mu (i\vec m)} \lambda^{I0}
- i m_a A_{\mu (i\vec m)} \lambda^{Ia}
\ea
where
\ba
(\hat D_\mu X^I)_{(i\vec m)} 
&=& \partial_\mu X^I_{(i\vec m)}
+ {f^{jk}}_i A_{\mu (k\vec n)} X^I_{(j,\vec m-\vec n)}\,,\\
A'_{\mu (i\vec m)} 
&=& -i m_a A_{\mu (i\vec m)}^a 
+ {f^{jk}}_i A_{\mu (j,\vec m-\vec n)(k\vec n)}\,.
\ea
We decompose this formula into the components
into the orthogonal spaces $\mathbf{R}^{7-d}$ and $\mathbf{R}^{d+1}$
by using the projector $P^{IJ}$ as
\ba\label{covd}
(D_\mu X^I)_{(i\vec m)}
=P^{IJ}(\hat D_\mu X^J)_{(i\vec m)} 
+\sum_A \lambda^{IA} (F_{\mu A})_{(i\vec m)}
\ea
where
\ba
(F_{\mu 0})_{(i\vec m)}
&=& \vec \pi_0\cdot(\hat D_\mu \vec X)_{(i\vec m)}
+ A'_{\mu (i,\vec m)}\nt
&=& \frac{1}{G^{00}} \hat D_\mu (\vec\lambda^0\cdot\vec X)_{(i\vm)}
- \frac{G^{0a}}{G^{00}} \hat D_\mu (\vec\pi_a\cdot \vec X)_{(i\vm)}
+ A'_{\mu (i\vec m)}\,,\\
(F_{\mu a})_{(i\vec m)}
&=& \hat D_\mu (\vec \pi_a\cdot\vec X)_{(i\vec m)}
- i m_a A_{\mu (i\vec m)}
\,.\label{Fma}
\ea
We will rewrite $\vec\pi_a\cdot \vec X$ as $A_a$ below,
since they play the role of gauge fields along the fiber $T^d$
as we mentioned. ${F_{\mu a}}$ will be regarded as
the field strength with one leg in $\mathcal{M}$ and the other in $T^d$.
${F_{\mu 0}}$ seems to be the field strength in a similar
sense with one leg in M-direction. 
However, the gauge field $A'_{\mu(i\vec m)}$ is an auxiliary field 
as we see below, and after it is integrated out, 
${F_{\mu 0}}$ will completely disappear from the action. 
In this sense, ${F_{\mu 0}}$ do not have any geometrical meaning.
We suspect, however, that 
it may give a hint to keep the trace of the compactification
of M-theory to type IIA superstring theory.

Finally, using eq.\,(\ref{covd}), the kinetic term for $X^I$ becomes
\ba
L_X
= -\frac12 \hat D_\mu X^I_{(i\vm)} P^{IJ} \hat D_\mu X^J_{(i,-\vm)}
- \frac12 G^{AB} {F_{\mu A (i\vm)}} {F_{\mu B (i,-\vm)}}\,.
\ea
Similarly, the kinetic term for $\Psi$ becomes
\ba
L_\Psi=\frac{i}{2}\bar\Psi_{(i\vm)}\Gamma^\mu\hat D_\mu \Psi_{(i,-\vm)}\,.
\ea

\paragraph{Chern-Simons term and integration of $A'$}

The Chern-Simons term is written as
\ba\label{eq:CS}
L_{CS}
&=& \frac12 \left(
 A'_{(i\vm)}\wedge d A_{(i,-\vm)} + A_{(i,-\vm)}\wedge d A'_{(i\vm)}
\right)\nt
&&-i f^{ijk}A'_{(i\vm)} \wedge A_{(j\vn)} \wedge A_{(k,-\vm-\vn)}\,,
\ea
or, up to the total derivative terms,
\ba
L_{CS}=\frac12 A'_{(i\vm)}\wedge F_{(i,-\vm)}+\mbox{(total derivative)}\,,
\ea
where
\ba
F_{\mu\nu(i\vm)}=\partial_\mu A_{\nu(i\vm)}-\partial_\nu A_{\mu(i\vm)}
+ {\ud f{jk}i} A_{\mu(j\vn)}A_{\nu(k,\vm-\vn)}\,.
\ea
Since the gauge field $A'$ shows up only in $L_{CS}$ and $L_X$,
one may algebraically integrate over it.
Variation of $A'$ gives the equation of motion gives
\ba
A'_{\mu (i,\vec m)}
&=& -\frac{1}{G^{00}}\hat D_\mu (\vec\lambda^0\cdot\vec X)_{(i\vm)}
+ \frac{G^{0a}}{G^{00}}\hat D_\mu {{A_a}_{(i\vm)}}
- \frac{G^{0a}}{G^{00}} ({F_{\mu a}})_{(i\vec m)}
\nt&&
- \frac{1}{2G^{00}} \epsilon_{\mu\nu\lambda} (F_{\nu\lambda})_{(i\vec m)}\,,
\ea
where $A_a:=\vec\pi_a\cdot\vec X$.
By putting back this value to the original action (\ref{eq:CS}),
\ba\label{S_kin1}
L_X+L_{CS}
&=&-\frac12\hat D_\mu X^I P^{IJ} \hat D_\mu X^J
-\frac1{4G^{00}} (F_{\nu\lambda})^2
-\frac12 \tilde G^{ab} {F_{\mu a}} {F_{\mu b}} \nt&&
-\frac{G^{0a}}{2G^{00}} \epsilon^{\mu\nu\lambda} {F_{\mu a}} F_{\nu\lambda}
+L_{td}\,,
\ea
where
\ba
\tilde G^{ab}&:=& G^{ab}-\frac{G^{a0}G^{b0}}{G^{00}}\,,\\
L_{td}&=&-\frac{1}{2G^{00}}\epsilon_{\mu\nu\lambda}\partial_\mu
\left[\left(
-i\hat D_\nu (\vec\lambda^0\cdot \vec X)
+\frac12\epsilon_{\nu\rho\sigma}F_{\rho\sigma}\right)
A_\lambda\right]\,.
\ea
Here we omit the indices $(i\vm)$ for simplicity.
Note that the redefinition of the metric $G^{ab}\to\tilde G^{ab}$ 
is very similar to that of T-duality transformation in M-direction.
The term $L_{td}$ is total derivative which does not vanish in the limit
$G^{0a}\to 0$.   Since we know that the total derivative terms
do not play any role for the case 
$G^{0a}=\vec\lambda^0\cdot\vec\lambda^a=0$,
we will neglect them in the following. 
In a sense, this is equivalent to redefine the BLG action,
\ba
S_{BLG}=\int d^3x\, (L_{BLG}-L_{td})\,,
\ea
where $L_{BLG}$ is the original BLG Lagrangian.
On the other hand, while the fourth term in eq.\,(\ref{S_kin1}) is also 
total derivative, we must {\em not} neglect it. This is because 
this term is proportional to $G^{0a}$ and becomes essential to 
understand the U-duality. For $d=1$ case, it becomes
the $\theta$ term of the super Yang-Mills action and it should be
involved in the S-duality transformation in the complex
coupling constant $\tau=C_0+ie^{-\phi}$.
We note that this is the term which does
not show up if we analyze the
Yang-Mills system with loop algebra symmetry (\ref{la}).

\paragraph{Kaluza-Klein mass by Higgs mechanism}
At this point, it is easy to understand how compactification occurs
after the Higgs mechanism. Note that in the definition of ${F_{\mu a}}$
(\ref{Fma}), we have a factor with $m_a$ in front of
$A_{\mu(i\vec m)}$.  In the language of D2-brane worldvolume,
it gives rise to the mass term
\ba 
-\frac12 g^{ab} m_a m_b A_{\mu (i\vec m)}
A^\mu_{(i,-\vec m)}\,,\quad\mbox{where}\quad
g^{ab}:=G^{00} \tilde G^{ab}, 
\ea
for
$A_{\mu(i\vec m)}$.  
We will also see that exactly the same mass term 
exists for all fields with index $\vec m$.  
It is natural to regard these terms as the Kaluza-Klein mass terms
for the compactification on a torus $T^d$.


In order to be more
explicit, we will use the T-dual picture \cite{Taylor:1996ik} 
in the following.
We identify the various fields with index $\vec m$ with the higher
$3+d$ dimensional fields by the identification
\ba\label{decomp1}
\Phi_{\vec m}(x) \rightarrow \tilde \Phi(x, y):=\sum_{\vec m} 
\Phi_{\vec m}(x) e^{i\vec m \vec y}
\ea
where $y^a\in [0,2\pi]$ ($a=1,\cdots,d$) are coordinates of $T^d$.
 ${F_{\mu a}}$ can be identified with
the field strength by
\ba
({{\tilde F}_{\mu a}})_i=\hat D_\mu \tilde A_{ai} 
-\frac{\partial}{\partial y^a} \tilde A_{\mu i}
\ea
where ${\tilde A_{a i}}(x,y):=\vec \pi_a \cdot \vec {\tilde X}_i (x,y)$.
The kinetic terms of gauge fields in eq.\,(\ref{S_kin1}) imply
that we have a metric in $\vec y$ direction as
\ba\label{Dpmet}
ds^2 =\eta_{\mu\nu} dx^\mu dx^\nu + g_{ab}\, dy^a dy^b\,,
\quad\mbox{where}\quad
g_{ab}:=(g^{ab})^{-1}\,.
\ea
When $\vec \lambda^A$ are all orthogonal, one may absorb the metric $g_{ab}$
in the rescaling of $y^a$ as 
$y'^a =(|\vec\lambda^0||\vec\lambda^a|)^{-1} y^a$.
Since $y^a$ has the radius $1$, 
$y'^a$ has the radius  $1/|\vec\lambda^0||\vec\lambda^a|$.
This is consistent with our previous analysis~\cite{HMS}.
In this scaling $y^a\to y'^a$, 
the kinetic terms for gauge fields in eq.\,(\ref{S_kin1}) become
\ba
-\frac{1}{4G^{00}} \left[(F_{\nu\lambda})^2 + 2 (F_{\mu a})^2 \right] \,,
\ea
which is also consistent with our previous study for $d=1$.

We note that the use of Kac-Moody algebra as the symmetry of the Kaluza-Klein
mode is not new.  See, for example, \cite{KM}.
Here the novelty is to use the Higgs mechanism to obtain
the Kaluza-Klein mass.

\paragraph{Worldvolume is a flat fiber bundle}
So far, since we put $A_\mu^a=0$ for the simplicity of the argument,
the worldvolume of D$p$-brane is the product space 
$\mathcal{M}\times T^d$.  In order to see the geometrical role
of $A_\mu^a$\,, let us keep it nonvanishing for a moment.
The covariant derivative (\ref{cvd}) get an extra term,
$
m_a A_\mu^a (x) X^I_{(i\vec m)}
$, which becomes
on $\mathcal{M}\times T^d$,
\ba
iA_\mu^a (x) \frac{\partial}{\partial y^a} \tilde X^I_i(x,y)\,.
\ea
$A_\mu^a$ turns out to be the gauge field for the
gauge transformation from those of  BLG:
\ba
\delta \tilde X^I_i(x,y) = i\gamma^a(x)  \frac{\partial}{\partial y^a}
\tilde X^I_i(x,y)\,.
\ea
The existence of the gauge coupling implies that the
worldvolume is not the direct product $\mathcal{M}\times T^d$
but a fiber bundle $Y$:
\[
\xymatrix{
{T}^d \ar[r] & Y \ar[d] \\
& \mathcal{M}
}
\]
where $T^d$ act as the translation of $y^a$.

The kinetic term for the connection comes from the Chern-Simons term:
\ba
L_{fiber}=\epsilon^{\mu\nu\lambda} {C_{\mu a}} \partial_\nu A_{\lambda}^a\,,
\quad
{C_{\mu a}}:=\sum_{\vec n} n_a A_{\mu (i\vec n) (i,-\vec n)}\,.
\ea
Since ${C_{\mu a}}$ does not appear in other place in the action,
its variation gives,
\ba
\partial_{[\mu } A_{\nu]}^a =0\,.
\ea
Therefore $Y$ must be a flat bundle as long as we start from 
BLG model.

There seems to be various possibilities to
relax this constraint to the curved background.
One naive guess is to replace $L_{fiber}$ to
\ba
L'_{fiber}=\epsilon^{\mu\nu\lambda} {C_{\mu a}} (\partial_\nu A_{\lambda}^a
-\frac{1}{2} F^{a(0)}_{\nu\lambda})\,,
\ea
for an appropriate classical background $F^{a(0)}_{\nu\lambda}$.

\paragraph{Interaction terms}
The compactification picture works as well in the interaction terms.
For the fermion interaction term $L_{int}$, we use
\ba
[X^{[I},X^{J]},\Psi]_{(i,-\vm)}
=- m_a\lambda^{[I0}\lambda^{J]a}\Psi_{(i,-\vm)}
+i{\ud f{jk}i}\lambda^{[I0}X^{J]}_{(j\vn)}\Psi_{(k,-\vm-\vn)}
\ea
and from eq.\,(\ref{eq:pi0}),
\ba \label{eq:expandX}
X^I
&=&P^{IJ}X^J+\lambda^{IA}(\vec\pi_A\cdot\vec X)\nt
&=&P^{IJ}X^J+\frac{1}{G^{00}}\lambda^{I0}(\vec\lambda^0\cdot\vec X)
+\left(-\frac{G^{0a}}{G^{00}}\lambda^{I0}+\lambda^{Ia}\right)A_a\,.
\ea
Then $L_{int}$ can be written as
\ba
L_{int}
&=& \frac{i}{4}\bar\Psi_{(i\vm)}(\Gamma_{IJ}\lambda^{I0}\lambda^{Ja})
\left(-m_a\Psi_{(i,-\vm)}
+i {\ud f{jk}i} {A_a}_{(j\vn)}\Psi_{(k,-\vm-\vn)}\right) \nt&&
+ \frac{i}{4}\bar\Psi_{(i\vm)}(\Gamma_{IJ}\lambda^{I0})\left(
i{\ud f{jk}i} P^{JK}X^K_{(j\vn)}\Psi_{(k,-\vm-\vn)}\right)\nt
&=& \int \frac{d^dy}{(2\pi)^d}\sqrt g\left(
\frac{i}{2}\tilde{\bar\Psi}\Gamma^a \hat D_a \tilde\Psi
+\frac{i\sqrt{G^{00}}}{2}\tilde{\bar\Psi}\Gamma_I [P^{IJ}\tilde X^J,\tilde \Psi]\right)\,,
\ea
where $g=\det\, g^{ab}$\,,
$\hat D_a\tilde\Psi:=\partial_a\tilde\Psi-i[\tilde A_a,\tilde\Psi]$
and
\ba
\Gamma^a:=\frac{i}{2}\Gamma_{IJ}\lambda^{I0}\lambda^{Ja}\,,\quad
\Gamma_J:=\frac{1}{2\sqrt{G^{00}}}\Gamma_{IJ}\lambda^{I0}\,,
\ea
which satisfy $\{\Gamma^a,\Gamma^b\}=g^{ab}$
and $\{\Gamma_I,\Gamma_J\}=\delta_{IJ}$.

On the other hand, the potential term for the boson $L_{pot}$ is 
the square of a 3-commutator:
\ba
[X^I, X^J, X^K]_{(i,\vec m)} = m_a \lambda^{[I0} \lambda^{Ja} X^{K]}_{(i,\vec m)}
+i {f^{jk}}_i \lambda^{[I0} X^J_{(j,\vec n)} X^{K]}_{(k,\vec m-\vn)}.
\ea
where the indices $I,J,K$ are antisymmetrized.
The square of the first term gives
\ba
\left(m_a \lambda^{[I0} \lambda^{Ja} X^{K]}_{(i,\vec m)}\right)^2
=6 g^{ab} m_a m_b X^I_{\vec m} P^{IJ}_{\vec m}  X^J_{-\vec m}\,,
\ea
where
\ba
P^{IJ}_{\vec m}&:=&\delta^{IJ}-
\frac{
 |\vec\lambda^0|^2\lambda_{\vec m}^I\lambda_{\vec m}^J
+|\lambda_{\vec m}|^2\vec\lambda^{I0}\vec\lambda^{J0}
-(\vec \lambda^0 \cdot \vec \lambda_{\vec m})
 (\lambda^{I0} \lambda_{\vec m}^J
+\lambda^{J0} \lambda_{\vec m}^I)}
{|\vec \lambda^0|^2 |\vec \lambda_{\vec m}|^2 
 -(\vec\lambda^0 \cdot \vec\lambda_{\vec m})^2}\,,\nt
\vec\lambda_{\vec m}&:=& m_a\vec \lambda^a\,,
\ea
which satisfy
\ba
P^{IJ}_{\vec m}  \lambda^{J0}=P^{IJ}_{\vec m} \lambda^J_{\vec m}=0\,,\quad
P_{\vec m}^2= P_{\vec m}\,.
\ea
The mixed term
\ba
\lambda^{[I0}\lambda_{\vec m}^J X^{K]}_{(i\vec m)}\cdot
{f^{jk}}_i \lambda^{[I0} X^J_{(j,\vec n)} X^{K]}_{(k,-\vec m-\vec n)}
\ea
vanishes and does not contribute to the action.
The commutator part is
\ba
(i{f^{jk}}_i \lambda^{[I0} X^J_{(j,\vec n)} X^{K]}_{(k,\vec m-\vn)})^2
=3\left(
G^{00} \langle [X^J, X^K]^2\rangle -2 \langle
[(\vec \lambda^0\cdot \vec X), X^I]^2\rangle
\right)
\ea
which is identical to the similar term in \cite{HIM}
and it produces the standard commutator terms.
Using eq.\,(\ref{eq:expandX}), these terms can be summarized
in the following compact form:
\ba
L_{pot}&=&\int\frac{d^dy}{(2\pi)^d}\sqrt g \left(
-\frac12 g^{ab}\hat D_a \tilde X^I P^{IJ} \hat D_b \tilde X^J
-\frac{1}{4G^{00}} g^{ac}g^{bd}\tilde F_{ab}\tilde F_{cd}
\right.\nt&&\quad\quad\quad\quad\quad\quad\left.
-
\frac{G^{00}}{4}[P^{IK}\tilde X^K,P^{JL}\tilde X^L]^2
\right)\,,
\ea
where $\hat D_a \tilde X^I=\partial_a \tilde X^I-i[\tilde A_a,\tilde X^I]$
and $\tilde F_{ab}=\partial_a \tilde A_b-\partial_b \tilde A_a-i[\tilde A_a,\tilde A_b]$.


\subsection{Summary}
\label{s:summary}

By collecting all the results in previous subsections,
the BLG action (\ref{S}) becomes
\ba \label{totalS}
L&=&L_A+L_{FF}+L_X+L_\Psi+L_{pot}+L_{int}+L_{td}\,,\\
L_A&=&-\frac{1}{4G^{00}}\int \frac{d^dy}{(2\pi)^d}\sqrt g\left(
\tilde F_{\mu\nu}^2
+2g^{ab}{\tilde F_{\mu a}}{\tilde F_{\mu b}}
+g^{ac}g^{bd}\tilde F_{ab}\tilde F_{cd}
\right)\,,\\
L_{FF}&=&-\frac{G^{0a}}{8G^{00}} \int \frac{d^dy}{(2\pi)^d}\sqrt g
\left(
4\epsilon^{\mu\nu\lambda} {\tilde F}_{\mu a} \tilde F_{\nu\lambda}\right)\,,\\
L_X&=&-\frac12\int \frac{d^dy}{(2\pi)^d}\sqrt g\left(
\hat D_\mu \tilde X^I P^{IJ} \hat D_\mu \tilde X^J
+g^{ab} \hat D_a \tilde X^I P^{IJ} \hat D_b \tilde X^J\right)\,,\\
L_\Psi&=&\frac{i}{2}\int \frac{d^dy}{(2\pi)^d}\sqrt g\,
\tilde{\bar\Psi}\left(
\Gamma^\mu \hat D_\mu +\Gamma^a \hat D_a\right)\tilde\Psi\,,\\
L_{pot}&=&-
\frac{G^{00}}4 \int \frac{d^dy}{(2\pi)^d}\sqrt g\, [P^{IK}\tilde X^K,P^{JL}\tilde X^L]^2\,,\\
L_{int}&=&\frac{i\sqrt{G^{00}}}2\int \frac{d^dy}{(2\pi)^d}\sqrt g\,
\tilde{\bar\Psi}\Gamma_I[P^{IJ}\tilde X^J,\tilde\Psi]\,.
\ea
It is easy to see that this is the standard 
D$p$-brane action ($p=d+2$) on $\mathcal{M} \times T^d$
with the metric (\ref{Dpmet}).
Interpretation and implications of this action are given
in the next section.

\section{Study of U-duality in L-BLG model}

\subsection{D3-branes case}
\label{sec:D3}

For $d=1$, if we write $\vec \lambda^0=\vec e^{\,0}$, 
$\vec \lambda^1=\tau_1 \vec e^{\,0}+\tau_2 \vec e^{\,1}$
(where $\vec e^{\,0} \cdot \vec e^{\,1} = 0$, $|\vec e^{\,0}| = |\vec e^{\,1}|$),
the action for the gauge field is given as
\ba
L_A+L_{FF}
&=&-\frac{1}{4G^{00}}\int \frac{dy}{2\pi}\sqrt g\,F^2
-\frac{G^{01}}{8G^{00}} \int \frac{dy}{2\pi}\,F\widetilde{F} \nt
&=& -\frac1{8\pi} \int dy \left(\frac{\tau_1}{2} F\widetilde{F}+\tau_2 F^2\right) 
\ea
where now $g=g^{11}$ and
\ba
F^2&=&\tilde F_{\mu\nu}^2+2g^{11}\tilde F_{\mu 1}\tilde F_{\mu 1}\,,\nt
F\widetilde{F}&=&(4\sqrt{g^{11}}\,\epsilon^{\mu\nu\lambda})
\tilde F_{\mu 1} \tilde F_{\nu\lambda}\,.
\ea
This shows that the action (\ref{totalS}) in this case is 
the standard D3-brane action with the $\theta$ term.

Under the $SL(2,\mathbf{Z})$ transformation
\ba
\left(
\begin{array}{c}
\vec \lambda^1\\ \vec \lambda^0
\end{array}
\right)
\rightarrow \left(
\begin{array}{cc}
a & b\\ c& d
\end{array}
\right)
\left(
\begin{array}{c}
\vec \lambda^1\\ \vec \lambda^0
\end{array}
\right)\,,
\ea
the moduli parameter $\tau$ is transformed as,
\ba
\tau \rightarrow \frac{a\tau +b}{c\tau +d}\,.
\ea
For $b=-c=1$, $a=d=0$, it gives rise to the standard
S-duality transformation $\tau \rightarrow -1/\tau$.
On the other hand, $a=d=1$, $b=n$ and $c=0$ gives 
the translation $\tau\rightarrow \tau+n$.

We do not claim that we have proven S-duality symmetry from
our model.  At the level of 3-algebra (\ref{dpa}--\ref{dpaf}), 
there is obvious asymmetry between $u_0, v^0$ and $u_1, v^1$.
Nevertheless, it is illuminating that the S-duality symmetry can be 
interpreted in so simple way.

On the other hand, the translation symmetry reduces 
to the automorphism of the 3-algebra (\ref{dpa}),
\ba
&& u_0 \rightarrow u_0 - n u_1 \,,\quad
u_1 \rightarrow u_1 \,,\nn\\
&& v^0 \rightarrow v^0 \,,\quad
v^1 \rightarrow v^1+n v^0 \,.
\ea
It is easy to see that the transformation
changes neither 3-algebra nor their  metric.
It induces the redefinition the ghost fields
as,
\ba
X^I= X^I_{u_0} u_0 +X^I_{u_1} u_1+\cdots
= X^I_{u_0} (u_0-n u_1) + (X^I_{u_1}+ n X^I_{u_0})\, u_1+\cdots\,.
\ea
It implies the transformation
$\vec \lambda^0\,\rightarrow\, \vec\lambda^0$\,, 
$\vec \lambda^1\,\rightarrow\, \vec\lambda^1+n\vec\lambda^0$\,.
Of course, at the classical level, there is no reason that the
parameter $n$ must be quantized.
It is interesting anyway that part of the duality transformation
comes from the automorphism of 3-algebra.

The T-duality transformation $\mathbf{Z}_2$
which interchanges D3- and D2-branes comes from the different identification of
component fields.  Namely, we have constructed 4-dimensional
field $\tilde X^I(x,y)$ from
the component fields $X^I_{(i\vec m)}(x)$ by Fourier series
(\ref{decomp1}).  One may instead interpret $X^I_{(i\vec m)}(x)$
as the 3-dimensional field and interpret $\vec m$ index as
describing open string mode which interpolate mirror images of
a  point in $T^1=\mathbf{R}/\mathbf{Z}$.  This is the standard
T-duality argument \cite{Taylor:1996ik}.

The relation between the coupling constant and the radius
in T-duality transformation is given as follows. 
Let us assume for a moment that $\vec\lambda^0\perp\vec\lambda^1$
for simplicity.
It is well known~\cite{MP} that 
putting a VEV $\vec X_{u_0}=\vec\lambda^0$ means the compactification of 
M-direction with the radius
\ba
R_0=|\vec\lambda^0|\, l_p^{3/2}\,,
\ea
where $l_p$ is 11-dimensional Planck length.
From the symmetry of $X_{u_0} \leftrightarrow X_{u_1}$, putting a VEV 
$\vec X_{u_1}=\vec\lambda^1$ must imply the compactification of 
another direction with the similar 
radius $\tilde R_1=|\vec\lambda^1|\, l_p^{3/2}$
before taking T-duality along $\vec\lambda^1$.
At this point, we have D2-brane worldvolume theory 
with string coupling
\ba
g_s=g_{YM}^2\,l_s=|\vec\lambda^0|^2\,l_s \,.
\ea
where $l_s$ is the string length, satisfying $l_p^3=g_sl_s^3$.
In \S\ref{sec:Dp}, we obtain D3-brane
since we compactify 
the $\vec\lambda^1$ direction with radius $\tilde R_1$
and simultaneously take T-duality for the same direction.
Thus the D3-brane is compactified on $S^1$ with the radius
\ba
R_1=\frac{l_s^2}{\tilde R_1}
=\frac{l_s^2}{|\vec\lambda^1|\sqrt{|\vec\lambda^0|^2\,l_s^4}}
=\frac{1}{|\vec\lambda^0||\vec\lambda^1|}\,,
\ea
and the string coupling for D3-brane worldvolume theory is
\ba
g_s'=g_s\frac{l_s}{\tilde R_1}=\frac{|\vec\lambda^0|}{|\vec\lambda^1|}\,.
\ea
This result is consistent with our result~\cite{HMS}, 
as we also discussed in \S\ref{sec:Dp}.

To summarize, the U-duality transformation
for $d=1$ case is
\ba
SL(2,\mathbf{Z}) \bowtie \mathbf{Z}_2\,,
\ea
where the first factor is described by the rotation of Higgs VEV's and
the second factor is described by the different representation
as the field theory.

\subsection{U-duality for $d>1$}
\label{sec:U-Dp}

We consider M-theory compactified on $T^{d+1}$ (where $d=p-2$).
This theory has U-duality group $E_{d+1}(\mathbf{Z})$ 
and scalars taking values in $E_{d+1}/H_{d+1}$ 
where $H_{d+1}$ is the maximal compact subgroup of $E_{d+1}$.
See, for example, \cite{Hull} for detail.
We call the space of these scalars `parameter space' in the following.

In this subsection, we compare the parameters obtained 
from L-BLG model with that in the parameter space.
We can extract various parameters on D$p$-brane from
the action obtained in \S\ref{s:summary}
which are all determined by the Higgs VEV's $\vec\lambda^A$.
The first one is the Yang-Mills coupling\,:
\ba
g_{YM}^2=\frac{(2\pi)^dG^{00}}{\sqrt g}\,,\quad
g:=\det\,g^{ab}\,.
\ea
Secondly, the metric 
\ba
g^{ab}=G^{00}G^{ab}-G^{0a}G^{0b}
\ea 
gives 
the moduli of the torus $T^d$.
Finally, $L_{FF}$ gives a generalization of $\theta$ term for $d=1$ case.
Since the $\theta$ term may be regarded as
the axion coupling, a natural generalization for general $d$ is
the R-R field $C_{(d-1)}$,
which appears in the D$p$-brane Lagrangian of string theory like as 
$C_{(d-1)}\wedge F\wedge F$.
Such term was discussed in the literature,  for example, in \cite{Hull}.

In our set-up in \S\ref{sec:Dp}, the existence of such coupling 
$C\wedge F\wedge F$ can be understood
as follows.  
There the compactification of the M-direction 
was determined by $\vec\lambda^0$ and 
we took T-duality on $T^d$ specified by 
$\{\vec\lambda^a\}=\{\vec\lambda^1,\cdots,\vec\lambda^d\}$.
If $G^{0a}=\vec\lambda^0\cdot\vec\lambda^a \neq 0$,
we obtain the non-zero $C_{(0)}$ field, 
after the compactification of M-direction and 
the T-duality transformation along only $y^a$.
After taking T-duality in the remaining $d-1$ directions on $T^d$ too, 
we 
obtain the nonzero $C_{(d-1)}$ field whose 
nonvanishing component is $C_{1\cdots\hat{a}\cdots d}$\,,
where the index with \,$\hat{}$\, should be erased.
This compontent of R-R field must interact with gauge fields on D-brane as
 $\epsilon^{\mu\nu\lambda 1\cdots d}C_{1\cdots\hat a\cdots 
d}F_{\mu\nu}F_{\lambda a}$. In our action 
(\ref{totalS}), $L_{FF}$ describes this coupling.
It determines the components of $C_{(d-1)}$ as
\ba\label{eq:Ca}
C_\ha:=C_{1\cdots\hat{a}\cdots d}
=\frac{1}{4(2\pi)^d(d-1)!}\frac{G^{0a}}{G^{00}}\frac{\sqrt g}{\sqrt{g^{aa}}}\,,
\ea
where no sum is taken on $a$.

The number of parameters thus obtained is
$1+\frac{d(d+1)}{2}+d =\frac{(d+1)(d+2)}{2}$
which coincides with the number of metric 
$G^{AB}=\vec \lambda^A\cdot \vec \lambda^B$.
As is $d=1$ case, it is natural to guess 
the $SL(d+1,\mathbf{Z})$ transformation
\ba \label{eq:SLdZ}
{\vec\lambda'^A}={\ud \Lambda A B}\,\vec\lambda^B\,,\quad
{\ud \Lambda A B} \in SL(d+1,\mathbf{Z})\,,
\ea
is related to the first factor of $SL(d+1,\mathbf{Z})\bowtie
O(d,d:\mathbf{Z})$
in U-duality transformation.
In appendix \ref{sec:SLdZ}, we derive the transformation law 
of these parameters explicitly.  They are less illuminative
compared with $d=1$ case,
however, since these parameters depends on $G^{AB}$ in a complicated way.
Since the number of the parameters is the same, it is straightforward
to obtain the inverse relation, 
$G^{AB}=G^{AB} (g_{YM}^2, g^{ab}, C_\ha)$. 
This combination transforms linearly under $SL(d+1,\mathbf{Z})$.
In this sense, it is possible to claim that $SL(d+1,\mathbf{Z})$ is
a part of the U-duality symmetry and $G^{AB}$ gives the parameter
which transforms covariantly under $SL(d+1,\mathbf{Z})$.
The closure of these parameters under $SL(d+1,\mathbf{Z})$
was discussed in the literature, for example, \cite{Hull}.

The parameters obtained from $\vec \lambda^A$, however,
do not describe the full parameter space to implement U-duality.
In the following, we compare it with the dimensions
of the parameter space.  As we see, for $d=1$, it correctly reproduces
the moduli.
The discrepancy of the number of
parameters starts from  $d>1$.  We will explain
some part of the missing parameters is given as the deformation
of 3-algebra (\ref{dpa}).


\paragraph{D3-brane $(d=1)$\,:}

It corresponds to M-theory compactified on $T^2$.
The parameter space in this case is 
$\bigl(SL(2)/U(1)\bigr)\times \bR$ 
which gives 3 scalars.
They correspond to $G^{00}$, $G^{01}$ and $g$,
in other words, $g_{YM}^2$, $C_{\hat 1}$ and $g^{11}$,
all of which appear in the D3-brane action (\ref{totalS}).

\paragraph{D4-branes $(d=2)$\,:}

It corresponds to M-theory compactified on $T^3$.
The parameter space in this case is 
$\bigl(SL(3)/SO(3)\bigr)\times \bigl(SL(2)/U(1)\bigr)$
which gives 7 parameters.
They correspond to $G^{ab}$, $B_{ab}$, $\Phi$ and $C_\ha$
which transform in the ${\bf 3} + {\bf 1} + {\bf 1} + {\bf 2}$ 
representations of $SL(2)$.
$\Phi$ is dilaton which satisfies
$e^{\Phi}=g_s=(2\pi)^{p-2}l_s^{p-3}g_{YM}^2$\,,
and $C_\ha$ is R-R 1-form (or $p-3$ form) field 
defined in eq.\,(\ref{eq:Ca}).

$B_{ab}$ is NS-NS 2-form field which we have not discussed so far.
As we commented in the footnote \ref{f:a}, such parameters
were introduced in section 5.2 of \cite{HMS}
as the deformation of the 3-algebra,
$[u_0,u_a,u_b]=B_{ab}T_{\vec 0}^0, \ \cdots$.
It describes the noncommutativity on the torus
along the line of \cite{HWW}.
We have not used this generalized algebra for the simplicity
of the computation but can be 
straightwardly included in the L-BLG model.
It is interesting that some part of moduli are described
as dynamical variable (``Higgs VEV") while the other part
comes from the modification of 3-algebra which underlies
the L-BLG model.


\paragraph{D5-branes $(d=3)$\,:}

It corresponds to M-theory compactified on $T^4$.
The parameter space in this case is $SL(5)/SO(5)$ 
which gives 14 parameters.
They correspond to 
$G^{ab}$, $B_{ab}$, $\Phi$, $C_\ha$ and $C_{\ha\hb\hc}$
which transform in the ${\bf 6} + {\bf 3} + {\bf 1} + {\bf 3} + {\bf 1}$
representations of $SL(3)$.

$C_{\ha\hb\hc}:=C_{1\cdots\ha\cdots\hb\cdots\hc\cdots d}$ 
is R-R 0-form (or $p-5$ form) field 
which causes the interaction like as
$\epsilon^{\mu\nu\lambda 1\cdots d}C_{\ha\hb\hc}F_{\mu\nu}F_{\lambda a}F_{bc}$ or
$\epsilon^{\mu\nu\lambda 1\cdots d}C_{\ha\hb\hc}F_{\mu a}F_{\nu b}F_{\lambda c}$\,.
In the context of 3-algebra, there is a room to include such coupling
\cite{HMS}. It is related to the 3-algebra associated with
Nambu-Poisson bracket.
As shown in~\cite{M5}, the worldvolume theory becomes not the
super Yang-Mills but instead described by self-dual 2-form field
which describes the M5-brane.
\footnote{In order to satisfy the fundamental identity, Nambu-Poisson
bracket must be equipped on a 3-dimensional manifold.
So, in this case, we must choose the specific $T^3$ where Nambu-Poisson 
bracket is defined from the whole compactified torus $T^4$.}
The precise statement on the moduli becomes obscure
in this sense.

To see U-duality, we must also consider the transformation of 
$B_{ab}$ and $C_{\ha\hb\hc}$. 
Especially, the interchange $B_{ab}\leftrightarrow C_\ha$ and
$C_{\ha\hb\hc}\leftrightarrow \Phi$ means S-duality.

\paragraph{D6-branes $(d=4)$\,:}
It corresponds to M-theory compactified on $T^5$.
The parameter space in this case is 
$SO(5,5)/\bigl(SO(5)\times SO(5)\bigr)$ 
which gives 25 scalars.
They correspond to 
$G^{ab}$, $B_{ab}$, $\Phi$, $C_{\hat a}$ and $C_{\hat a\hat b\hat c}$
which transform in the 
${\bf 10} + {\bf 6} + {\bf 1} + {\bf 4} + {\bf 4}$ 
representation of $SL(4)$.
To see U-duality, we must also consider the transformation of 
$B_{ab}$ and $C_{\ha\hb\hc}$.

\paragraph{D7-branes $(d=5)$\,:}

It corresponds to M-theory compactified on $T^6$.
The parameter space in this case is $E_6/USp(8)$
which gives 42 scalars.
They correspond to 
$G^{ab}$, $B_{ab}$, $\Phi$, $C_{\hat a}$, $C_{\hat a\hat b\hat c}$
and $C_{\ha\hb\hc\hd\he}$
which transform in the 
${\bf 15} + {\bf 10} + {\bf 1} + {\bf 5} + {\bf 10} + {\bf 1}$ 
representations of $SL(5)$.

$C_{\ha\hb\hc\he\hf}$ is R-R 0-form (or $p-7$ form) field 
which causes the interaction like as
$\epsilon^{\mu\nu\lambda 1\cdots d}C_{\ha\hb\hc\he\hf}F_{\mu\nu}F_{\lambda a}F_{bc}F_{ef}$ and so on.
Note that $C_\ha$ in this case must be the self-dual 4-form field.

To see U-duality, we must also consider the transformation of 
$B_{ab}$, $C_{\ha\hb\hc}$ and $C_{\ha\hb\hc\he\hf}$. 
Especially, the interchange $B_{ab}\leftrightarrow C_{\ha\hb\hc}$ and
$C_{\ha\hb\hc\he\hf}\leftrightarrow \Phi$ means S-duality.
However we don't know the way to introduce the field $C_{\ha\hb\hc\he\hf}$
at this moment in time, so this discussion may be difficult.

\paragraph{D8-branes $(d=6)$\,:}

It corresponds to M-theory compactified on $T^7$.
The parameter space in this case is $E_7/SU(8)$ 
which gives 70 scalars.
They correspond to 
$G^{ab}$, $B_{ab}$, $\Phi$, $C_{\hat a}$, $C_{\hat a\hat b\hat c}$
and $C_{\ha\hb\hc\he\hf}$
which transform in the 
${\bf 21} + {\bf 15} + {\bf 1} + {\bf 6} + {\bf 20} + {\bf 6}$
representations of $SL(6)$,
plus one additional scalar $B_{abcefg}$ which is the dual of NS-NS
2-form $*B_{(2)}$.
To see U-duality, we must consider the transformation of all these fields.

\paragraph{D9-branes $(d=7)$\,:}

It corresponds to M-theory compactified on $T^8$. 
The parameter space in this case is $E_8/SO(16)$ 
which gives 128 scalars.
They correspond to $G^{ab}$, $B_{ab}$, $\Phi$, $C_{\hat a}$, 
$C_{\hat a\hat b\hat c}$, $C_{\ha\hb\hc\he\hf}$ and 
$C_{\ha\hb\hc\he\hf\hg\hh}$
which transform in the 
${\bf 28} + {\bf 21} + {\bf 1} + {\bf 7} + {\bf 35} + {\bf 21} + {\bf 1}$
representations of $SL(7)$,
plus 14 additional scalars $B_{abcefg}$ and $C_{\mu a}$\,.
This $C_{\mu a}$ is R-R 2-form field which has legs belong to 
one of worldvolume coordinates $x^\mu$ and one of torus coordinates $y^a$.

To see U-duality, we must consider the transformation of all these fields.
However we don't know the way to introduce the field $C_{\ha\hb\hc\he\hf}$
and $C_{\ha\hb\hc\he\hf\hg\hh}$
at this moment in time, so this discussion may be very difficult.

\section{Conclusion and Discussion}

In this paper, we have presented a detailed derivation
of D$p$-brane action from BLG model.
The VEV's of ghost fields $\vec \lambda^A$ give 
the moduli of torus $T^d$ ($d=p-2$) $g^{ab}$,  
the coupling constants $g_{YM}$ of super Yang-Mills 
and the R-R $(p-3)$-form field $C_{\hat a}$ 
through the `metric' $G^{AB}=\vec\lambda^A\cdot\vec\lambda^B$.
For D3-branes ($d=1$), the parameters thus obtained are enough
to realize full  Montonen-Olive duality group 
$SL(2,\mathbf{Z})$ through the linear transformation
on $\vec\lambda^A$.
Moreover, some part of the symmetry 
is actually the automorphism of 3-algebra.
For higher dimensional case $d>1$ (D$p$-branes
with $p>3$), these parameters are enough to implement
a subgroup of U-duality transformation, $SL(d+1,\mathbf{Z})$,
which acts linearly on $\vec\lambda^A$.
The transformations of various parameters can be determined
through the linear transformation of the metric $G^{AB}$.
In order to realize the full U-duality group, however, they are not enough.
We argue that one of the missed parameters, NS-NS 2-form background,
can be introduced through the deformation of
the 3-algebra.  For higher $d$, we need extra R-R background
which we could not succeed to explain in the context of L-BLG models
so far. One possibility may be to use the coupling constants of Nambu-Poisson
bracket which gives rise to self-dual 2-form field on the worldvolume
instead of super Yang-Mills.


There are a few directions for the futher
development from current work.
One direction is to understand the higher $d$ case in more detail.
For higher $d$, we have to think more carefully
on the fundamental degree of freedom. In some cases,
the gauge theory should be replaced by 2-form fields,
and sometimes by strings.  We hope that the BLG description
of M5-brane~\cite{M5} gives an essential hint.

It is also interesting to derive the  U-duality symmetry from ABJM model.
While some work have been done in \cite{HTT} for D3-brane, it may be
interesting how to incorporate the loop algebras in ABJM context
which would help us to go beyond D3.
As we explained here, the loop algebra is suitable symmetry to describe
the Kaluza-Klein modes.

Another interesting direction is to describe the curved background
or D-branes from L-BLG model.  As we already explained in the text,
as long as we start from BLG model, we arrive at a flat background.
This is natural since we have maximal supersymmetry.
If, however, one modifies the action slightly (a naive discussion
is given in the text),  there is more room to incorporate
various degree of freedom.  Such modification of the model
seems essential to understand various M-brane dynamics.

\subsection*{Acknowledgment}

We would like to thank Pei-Ming Ho for sharing his excellent
insights through the collaboration at some stages.
Y. M. is partially supported by KAKENHI (\#20540253) from MEXT, Japan.
S. S. and T. K. are partially supported by Global COE Program 
``the Physical Sciences Frontier'', MEXT, Japan.

\appendix
\section{$SL(d+1,\mathbf{Z})$ transformations on D$p$-branes}
\label{sec:SLdZ}
In this appendix, we compute the transformation law
for the moduli parameters under $SL(d+1,\mathbf{Z})$
transformation (\ref{eq:SLdZ}).
$SL(d+1,\mathbf{Z})$ is generated by
the following two kinds of $(d+1)\times (d+1)$ matrices:
\ba
S(i,j)&:&\begin{cases}
&{\ud \Lambda A B}=\delta^A_B \quad(\mbox{for~}A,B\neq i,j)\,,\nn\\[+3pt]
&\begin{pmatrix}
{\ud \Lambda i i}&{\ud \Lambda i j}\\{\ud \Lambda j i}&{\ud \Lambda j j}
\end{pmatrix}
=\begin{pmatrix}0&1\\-1&0\end{pmatrix}\,.
\end{cases}\\
T(i,j;n)&:&\begin{cases}
&{\ud \Lambda A B}=\delta^A_B \quad(\mbox{for~}A,B\neq i,j)\,,\nn\\[+3pt]
&\begin{pmatrix}
{\ud \Lambda i i}&{\ud \Lambda i j}\\{\ud \Lambda j i}&{\ud \Lambda j j}
\end{pmatrix}
=\begin{pmatrix}1&0\\n&1\end{pmatrix}\,.
\end{cases}
\ea
where $i,j=0,1,\cdots, d$ ($i<j$) and $n\in \mathbf{Z}$.
Obviously, $S(i,j)$ is a generalization of S-duality transformation
and $T(i,j;n)$ is the generalization of translation generator.




\paragraph{(I) $\Lambda=S(0,i)$ $(i\neq 0)$\,:}

This transformation interchanges $\vec\lambda^0$ 
and $\vec\lambda^i$\,, {\em i.e.} M-direction and one of the torus directions.
It is a generalization of S-duality transformation for $d=1$ case.
$G^{0A}$ and $g^{ab}$ are transformed as
\ba
&&G'^{00}=G^{ii}\,,\quad
G'^{0i}=-G^{0i}\,,\quad
G'^{0a}=G^{ia}\,,\nt
&&g'^{ii}=g^{ii},\quad
g'^{ia}=-(G^{ii}G^{0a}-G^{i0}G^{ia})\,,\quad
g'^{ab}=G^{ii}G^{ab}-G^{ia}G^{ib}\,,
\ea
for $a,b\neq 0,i$. In the simple case of $G^{0a}=G^{0i}=G^{ia}=0$, 
\ba
g_{YM}^2=\sqrt{\frac{G^{00}}{G^{ii}}}\,\frac{(2\pi)^d}{(G^{00})^{(d-1)/2}}\,
\frac{1}{\sqrt{\hat G^i}}
\quad\to\quad
g_{YM}^{'2}=\sqrt{\frac{G^{ii}}{G^{00}}}\,\frac{(2\pi)^d}{(G^{ii})^{(d-1)/2}}\,
\frac{1}{\sqrt{\hat G^i}}
\,,
\ea
where ${\hat G^i}$ is the minor determinant of $G^{ab}$ excluding the $i$'th 
row and column. On the other hand, $C_\ha$ remains zero in this simple case.

\paragraph{(II) $\Lambda=T(0,i;n)$ $(i\neq 0)$\,:}

This transformation shifts the direction as 
$\vec\lambda^0\to \vec\lambda^0$ and 
$\vec\lambda^i\to \vec\lambda^i+n\vec\lambda^0$,
and should be a generalization of T-duality transformation.
$G^{0A}$ and $g^{ab}$ are transformed as
\ba
&&G'^{00}=G^{00}\,,\quad
G'^{0i}=G^{0i}+nG^{00}\,,\quad
G'^{0a}=G^{0a}\,,\nt
&&g'^{ii}=g^{ii}\,,\quad
g'^{ia}=g^{ia}\,,\quad
g'^{ab}=g^{ab}\,,
\ea
for $a,b\neq 0,i$.
So the coupling constant $g_{YM}^2$ is invariant under this transformation.
On the other hand, one component of R-R field $C_{(d-1)}$ is shifted as in the D3-branes case, 
\ba
C_{\hat i} ~~\to~~
C_{\hat i}' \,=\, C_{\hat i}
+\frac{n}{4(2\pi)^d(d-1)!}\frac{\sqrt{g}}{\sqrt{g^{ii}}}\,,
\ea
while all the other components remain the same.

\paragraph{(III) $\Lambda=S(i,j)$ $(i,j\neq 0)$\,:}

This transformation interchanges  $\vec\lambda^i$ and $\vec\lambda^j$ and
should make no physical change. In fact,
\ba
&&G'^{00}=G^{00}\,,\quad
G'^{0i}=G^{0j}\,,\quad
G'^{0j}=-G^{0i}\,,\nt
&&g'^{ii}=g^{jj}\,,\quad
g'^{ij}=-g^{ji}\,,\quad
g'^{ji}=-g^{ij}\,,\quad
g'^{jj}=g^{ii}\,,
\ea
and other $G^{0a}$ and $g^{ab}$ remain the same.
The coupling constant $g_{YM}^2$ is invariant under this transformation.
The components of $C_{(d-1)}$ is shuffled by the interchange of
the basis $\{\vec \lambda^a\}$, but this doesn't mean any physical changes.

\paragraph{(IV) $\Lambda=T(i,j;n)$ $(i,j\neq 0)$\,:}

This transformation shifts the torus direction as
$\vec\lambda^i\to \vec\lambda^i$ and 
$\vec\lambda^j\to \vec\lambda^j+n\vec\lambda^i$.
In this case, $G^{0A}$ and $g^{ab}$ are transformed as
\ba
&&G'^{00}=G^{00}\,,\quad
G'^{0j}=G^{0j}+nG^{0i}\,,\quad
G'^{0a}=G^{0a}\,,\nt
&&g'^{jj}=g^{jj}+2ng^{ji}+n^2g^{ii}\,,\quad
g'^{ja}=g^{ja}+ng^{ia}\,,\quad
g'^{ab}=g^{ab}\,,
\ea
for $a,b\neq 0,j$. 
Since $\sqrt g$ (or the volume of $T^d$) remains the same,
$g_{YM}^2$ is invariant under this transformation.
The components of $C_{(d-1)}$, just as in the case of $S(i,j)$, 
is effected by the transformation of the basis $\{\vec \lambda^a\}$, 
but it is not physically meaningful.

\paragraph{}



As we discussed in \S\ref{sec:U-Dp}, 
the transformation laws are somewhat complicated,
since the parameters $g_{YM}^2$ and $C_{\hat a}$ depends on 
$G^{00}$ and $G^{0a}$ in complicated way.
So if we want to see concisely the correspondence 
between subgroup of U-duality $SL(d+1,\mathbf{Z})$ 
and transformation of VEV's (\ref{eq:SLdZ}),
we must notice the transformation of 
$G^{AB}=\vec\lambda^A\cdot \vec\lambda^B$\,.
In fact, $G^{AB}$ is the linear realization of 
$SL(d+1,\mathbf{Z})$ transformation (\ref{eq:SLdZ}), 
and the parameters $g_{YM}^2$ and $C_{\ha}$ transform complexly 
through this covariant transformation of $G^{AB}$.

\end{document}